\documentclass[superscriptaddress]{revtex4}
 \usepackage{amsmath,lscape,epsfig}

 \def\beq{\begin{equation}}
 \def\eeq{\end{equation}}
 \def\beqa{\begin{eqnarray}}
 \def\eeqa{\end{eqnarray}}
 \def\ban{\begin{eqnarray*}}
 \def\ean{\end{eqnarray*}}
 \def\bi{\begin{itemize}}
 \def\ei{\end{itemize}}

\begin{document}
 
 \title{Metastability of hadronic compact stars}
 \author{Ignazio Bombaci}
 \affiliation{Dipartimento di Fisica ``Enrico Fermi'', Universit\`a di Pisa,
and INFN Sezione di Pisa,   
 Largo Bruno Pontecorvo 3, I-56127 Pisa, Italy }
 \author{Prafulla K. Panda}
 \affiliation{Indian Association for the Cultivation of Sciences, 
 Jadavpur, Kolkata-700 032, India}
 \affiliation{Centro de F\'{\i}sica Te\'orica, Department of Physics,
 University of Coimbra, 3004-516 Coimbra, Portugal}
 \author{Constan\c{c}a Provid\^encia}
 \affiliation{Centro de F\'{\i}sica Te\'orica, Department of Physics,
 University of Coimbra, 3004-516 Coimbra, Portugal}
 \author{Isaac Vida\~na}
 \affiliation{Departament d'Estructura i Constituents de la Mat\`eria. 
 Universitat de Barcelona, Avda. Diagonal 647, E-08028 Barcelona, Spain}

\begin{abstract}
Pure hadronic compact stars, above a threshold value of their gravitational mass 
(central pressure), are metastable to the conversion to quark stars (hybrid or strange stars).  
In this paper, we present a systematic study of the metastability of pure hadronic compact stars 
using different relativistic models for the equation of state (EoS).   
In particular, we compare results for the quark-meson coupling (QMC) model  with those for the Glendenning--Moszkowski parametrization of the non-linear Walecka model (NLWM). 
For QMC model, we find  large values ($M_{cr} = 1.6$ -- $1.9 M_\odot$) for the critical mass of 
the hadronic star sequence and we find that the formation of a quark star is only possible with 
a soft quark matter EoS.  
For the Glendenning--Moszkowski parametrization of the NLWM, we explore the effect of different hyperon couplings on the critical mass and on the stellar conversion energy. 
We find that increasing the value of the hyperon coupling constants shifts the bulk transition point 
for quark deconfinement to higher densities, increases the stellar metastability threshold mass 
and the value of the critical mass, and thus makes the formation of quark stars less likely.  
For the  largest values of the hyperon  couplings we find a critical mass which may be as 
high as 1.9 - 2.1 $M_\odot$. These stellar configurations, which contain a large central 
hyperon fraction ($f_{Y,cr} \sim 30 \%$),  would be able to describe highly-massive compact stars, 
such as the one associated to the millisecond pulsars PSR~B1516+02B with a mass 
$M = 1.94^{+ 0.17}_{- 0.19} M_{\odot}$.  
\end{abstract}

 \maketitle
 
 \vspace{0.50cm}
 PACS number(s): {97.60.s, 97.60.Jd, 26.60.Dd, 26.60.Kp}  
 \vspace{0.50cm}

 \section{Introduction}
 The nucleation of quark matter in neutron stars has been studied by many authors, due to
 its potential connection with explosive astrophysical events such as supernovae and 
 gamma ray burst. Some of the earlier studies on quark matter nucleation (see e.g.,
 \cite{ho92,ho94,ol94} and references therein) dealt with thermal
 nucleation in hot and dense hadronic matter. In these studies, it was found that 
 the prompt formation of a critical size drop of quark matter via thermal activation 
 is possible above a temperature of about $2-3$ MeV. As a consequence, it was inferred
 that pure hadronic stars are converted to quark stars (hybrid stars (HyS) or strange stars (SS))
 within the first seconds after their birth. However, neutrino trapping in the protoneutron
 star phase strongly precludes the formation of a quark phase \cite{prak97,lu98,be99,vi05}. 
 Then, it is possible that the compact star survives the early stages of its evolution as a pure 
 hadronic star. In this case, the nucleation of quark matter would be triggered by quantum 
 fluctuations in degenerate ($T=0$) neutrino free hadronic matter 
 \cite{gr98,iida97,iida98,be03,bo04,drago04,lug05,blv07}. 
 
 Quantum fluctuations could form, in principle, a drop of $\beta$-stable quark matter (hereafter 
 the $Q^{\beta}$ phase). However, this process is strongly suppressed with respect to the formation 
 of a non $\beta$-stable drop by a factor $\sim G^{2N/3}_{Fermi}$, where $N\sim 100-1000$ is 
 the number of quarks in a critical-size quark drop. This is so because the formation of a 
 $\beta$-stable drop would involve the almost simultaneous conversion of $\sim N/3$ up ($u$) and 
 down ($d$) quarks into strange ($s$) quarks. 
 Alternatively, quantum fluctuations can form a non $\beta$-stable drop (hereafter the $Q^*$ phase), 
 in which the flavor content of the quark phase is equal to that of the $\beta$-stable hadronic 
 phase at the same pressure \cite{iida97,iida98,bo04}. 
 Since no flavor conversion is involved, there are no suppressing Fermi factors, and a $Q^*$ 
 drop can be nucleated much more easily. 
 Once a critical-size $Q^*$ drop is formed, the weak interactions will have enough time to act, 
 changing the quark flavor fraction of the deconfined droplet to lower its energy, and a drop of 
 the $Q^\beta$ phase is formed. 
 This first seed of quark matter will trigger the conversion \cite{oli87,hbp91,grb} of the pure hadronic 
 star to a hybrid star or to a strange star (depending on the details of the 
 equation of state for quark matter used to model the phase transition). 
 The stellar conversion process liberates a total energy of the order of $10^{53}$ erg \cite{grb}. 
 
 When finite-size effects at the interface between the quark and hadron phases are taken into 
 account, it is necessary to have an overpressure $\Delta P = P - P_0> 0$ with respect to the 
 bulk transition point $P_0$, to create a drop of deconfined quark matter. 
 As a consequence, pure hadronic stars with values of the central pressure larger than 
 $P_0$ are metastable to the decay (conversion) to hybrid stars or to strange 
 stars \cite{be03,bo04,drago04,lug05,blv07}. The mean lifetime of the metastable stellar configuration 
 is related to the time needed to nucleate the first drop of quark matter in the stellar center and 
 depends dramatically on the value of the stellar central 
pressure \cite{be03,bo04,drago04,lug05,blv07}. 
 
The possibility of having in nature both metastable hadronic stars and stable quark stars, 
has led the authors of ref. \cite{bo04} to extend the concept of limiting mass of a 
``neutron star'' with respect to the {\it classical} one  introduced by 
Oppenheimer and Volkoff \cite{ov39}.    
Since metastable HS with a ``short'' {\it mean-life time} are very unlikely to be observed,  
the extended concept of limiting  mass  has been  introduced in view of the comparison 
with  the values of the mass of compact stars deduced from direct astrophysical 
observation (see sect. 3.1 of ref. \cite{bo04} for the definition of the {\it limiting mass}, 
$M_{lim}$, of compact stars in the case of metastable pure  hadronic stars).  

As it is well known, neutron star mass measurements give one of the most stringent test 
on the overall {\it stiffness} of dense matter EoS.    
Recent measurements of Post Keplerian orbital parameters in relativistic binary stellar systems (containing millisecond pulsars) give strong evidence for the existence of highly-massive 
``neutron stars''.  For example, the compact star associated to the millisecond pulsar 
PSR~B1516+02B in the Globular Cluster NGC~5904 (M5) has a mass  
$M = 1.94^{+ 0.17}_{- 0.19} M_{\odot}$  (1 $\sigma$) \cite{frei07a}.
In the case of PSR~J1748-2021B, a millisecond pulsar in the Globular Cluster NGC~6440, 
the measured mass is  $M = 2.74^{+ 0.41}_{- 0.51} M_{\odot}$  (2 $\sigma$) \cite{frei07b}. 
These measurements challenge most of the existing models for dense matter EoS.  

 In this work, we carry out a systematic study of the properties of metastable hadronic 
compact stars obtained within different relativistic mean-field models for the equation of state (EoS) of  hadronic matter. In particular, we compare the predictions of the Quark-Meson Coupling (QMC) 
model \cite{guichon,ST} with those of the non-linear Walecka model (NLWM) \cite{qhd}   parametrizations given by  Glendenning--Moszkowski (GM) \cite{gm91}.

For the quark phase we have adopted a phenomenological EOS \cite{farhi} which is 
based on the MIT bag model for hadrons. The parameters here are: the mass
$m_s$ of the strange quark, the so-called pressure of the vacuum $B$ (bag
constant) and the QCD structure constant $\alpha_s$. For all the quark
matter model used in the present work, we take $m_u = m_d =0$, 
$m_s = 150$~MeV and $\alpha_s = 0$. 

In the QMC model quark degrees of freedom are explicitly taken into account: 
baryons are described as a system of non-overlapping MIT bags which interact through the 
effective scalar and vector mean fields.  The coupling constants are defined at the quark level. 
An attractive aspect of the model is that different phases of hadronic matter, from very low to 
very high baryon densities, can be described within the same underlying model, namely the 
MIT bag model: matter at low densities is a system of nucleons interacting through meson fields, 
with quarks and gluons confined within MIT bag; at very high density one expects that baryons 
and mesons dissolve and the entire system of quarks and gluons becomes
confined within a single, big MIT bag.   

In the case of the Glendenning--Moszkowski EoS \cite{gm91}, we have  paid special attention 
to the role played by the hyperon-meson couplings. In fact, all previous works on metastable 
hadronic stars \cite{be03,bo04,drago04,lug05,blv07} have uniquely considered the case 
of ``low'' values for these quantities ($x_\sigma = 0.6$ for the ratio between the 
hyperon--$\sigma$ meson to nucleon--$\sigma$ meson coupling).   
As it is well known, larger values of the hyperon-meson couplings (constrained by the empirical 
binding energy of the $\Lambda$ particle in nuclear matter) make the EoS stiffer and increase the value of the Oppenheimer--Volkoff mass for the hadronic stellar sequence \cite{gm91}. 
In addition, as we demonstrate in the present work, increasing the values of the hyperon-meson couplings shifts the bulk transition point for quark deconfinement to higher densities and 
increments the value of the {\it critical mass} $M_{cr}$ (see ref.\cite{be03,bo04,drago04} and  
Section \ref{sec:quantum} for the explicit definition of this quantity) for the hadronic stellar 
sequence. Thus our study is relevant in connections with the recent measurements of  
highly-massive ``neutron stars'' mentioned above.  

 A brief review of the NLW and QMC models is given in Section \ref{sec:formalism}.
 The quantum nucleation of a quark matter drop inside hadronic matter is briefly reviewed 
in Section \ref{sec:quantum}. Our main results are presented in Section \ref{sec:results}, 
whereas the main conclusions are given in Section \ref{sec:conclusions}

 \section{The formalism}
 \label{sec:formalism}
 In the present section we review the models used in this work, namely the GM
 parametrizations \cite{gm91} of the NLWM and the quark-meson coupling (QMC)
 model including hyperons.

 \subsection{The non-linear Walecka model}
  The Lagrangian density, including the baryonic octet, in terms of the scalar
  $\sigma$, the vector-isoscalar $\omega_\mu$ and the vector-isovector $\vec \rho_\mu$
  meson fields reads (see {\it e.g.} \cite{prak97,glen00,mp03}) 
 
 \begin{equation}
 {\cal L}={\cal L}_{hadrons}+{\cal L}_{leptons}
 \end{equation}
 
 where the hadronic contribution is
 
 \begin{equation}
 {\cal L}_{hadrons}={\cal L}_{baryons}+{\cal L}_{mesons}
 \end{equation}
 
 with
 
 \begin{equation}
 {\cal L}_{baryons}=\sum_{\mbox{baryons}} \bar \psi\left[\gamma^\mu D_\mu -M^*_B\right]\psi,
 \end{equation}
 
 where
 \begin{equation}
 D_\mu=i\partial_{\mu}
 -g_{\omega B} \omega_{\mu}-{g_{\rho B}} \vec{t_B} \cdot \vec{\rho}_\mu, 
 \end{equation}
 and 
 $M^*_B=M_B-g_{\sigma B} \sigma.$ The quantity $\vec{t_B}$ designates the isospin of
 baryon $B$.
 The mesonic contribution reads 
 \begin{equation}
 {\cal L}_{mesons}={\cal L}_{\sigma}+{\cal L}_{\omega}+ {\cal L}_{\rho},
 \end{equation}
 with
 \begin{equation}
     {\cal L}_\sigma=\frac{1}{2}(\partial_{\mu}\sigma\partial^{\mu}\sigma
     -m_{\sigma}^2 \sigma^2)+ \frac{1}{3!} \kappa \sigma^3+ \frac{1}{4!} \lambda
     \sigma^4,
 \end{equation}
 \begin{equation}
     {\cal L}_{\omega}=-\frac{1}{4}\Omega_{\mu\nu}\Omega^{\mu\nu}+\frac{1}{2}
     m_{\omega}^2 \omega_{\mu}\omega^{\mu}, \qquad \Omega_{\mu\nu}=\partial_{\mu}\omega_{\nu}-\partial_{\nu}\omega_{\mu},
 \end{equation}
 \begin{equation}
     {\cal L}_{\rho}=
    { -\frac{1}{4}\vec B_{\mu\nu}\cdot\vec B^{\mu\nu}}+\frac{1}{2}
     m_\rho^2 \vec \rho_{\mu}\cdot \vec \rho^{\mu}, \quad \vec B_{\mu\nu}=\partial_{\mu}\vec \rho_{\nu}-\partial_{\nu} \vec \rho_{\mu}
       - g_\rho (\vec \rho_\mu \times \vec \rho_\nu)
 \end{equation}
 For the lepton contribution we take
 \begin{equation}
 {\cal L}_{leptons}=\sum_{\mbox{leptons}} \bar \psi_l \left(i \gamma_\mu \partial^{\mu}-
 m_l\right)\psi_l,
 \end{equation}
 where the sum is over electrons and muons.
 In uniform matter, we get for the baryon Fermi energy
 $
       \epsilon_{FB}=g_{\omega B} \omega_0+ g_{\rho B} t_{3B} \rho_{03} + \sqrt{k_{FB}^2+{M^*_B}^2},
 $
 with the baryon effective mass
    $M^*_B=M-g_{\sigma B}\sigma.$
 
 We will use the GM1 and GM3 parametrizations of NLWM \cite{gm91} both fitted to
 the bulk properties of
    nuclear matter: for GM1 (GM3) the compressibility is 300 (240) MeV and the
    effective mass at saturation
    $M^{*} = 0.7\, M$ ($M^{*} = 0.78\, M$). 
 The inclusion of hyperons involves new couplings, the hyperon-nucleon couplings:
 $g_{\sigma B}=x_{\sigma B}~ g_{\sigma},~~g_{\omega B}=x_{\omega B}~ g_{\omega},~~g_{\rho B}=x_{\rho B}~ g_{\rho}$.
 For nucleons we take
  $x_{\sigma B}$, $x_{\omega B}$, $x_{\rho B} = 1$
 and for hyperons we will consider the couplings proposed by Glendenning and Moszkowski \cite{gm91}.
 They have considered the binding energy of the $\Lambda$ in nuclear
 matter, $B_\Lambda$, 
 \begin{equation}
 \left(\frac{B_\Lambda}{A}\right)=-28 \mbox{ MeV}= x_{\omega} \, g_{\omega}\, \omega_0-x_{\sigma}\, g_{\sigma} \sigma
 \end{equation}
 to establish a relation between $x_{\sigma}$ and $x_{\omega}$. 
 Moreover, known neutron star masses restrict $x_{\sigma}$ to the range $0.6-0.8$. We
 will take $x_{\rho}=x_{\sigma}$ and will consider $x_{\sigma}=0.6,\,0.7,\,0.8$.

 \subsection{The quark-meson coupling model for hadronic matter}
 
 In the QMC model, the nucleon in nuclear medium is assumed to be a
 static spherical MIT bag in which quarks interact with the scalar and
 vector fields, $\sigma$, $\omega$ and $\rho$ and these
 fields are treated as classical fields in the mean field
 approximation \cite{guichon,ST}. 
 The quark field, $\psi_q(x)$, inside the bag then 
 satisfies the equation of motion: 
 \begin{equation}
 \left[i\,\rlap{/}\partial-(m_q^0-g_\sigma^q\, \sigma)
 -g_\omega^q\, \omega\,\gamma^0 + \frac{1}{2} g^q_\rho \tau_z \rho_{03}\right]
 \,\psi_q(x)=0\ , \quad q=u,d,s,
 \label{eq-motion}
 \end{equation}
 where $m_q^0$ is the current quark mass, and $g_\sigma^q$,
 $g_\omega^q$ and $g_\rho^q$ and denote the quark-meson coupling constants. 
 The normalized ground state for a quark in the bag is given
 by 
 \begin{equation}
 \psi_q({\bf r}, t) = {\cal N}_q \exp \left(-i\epsilon_q t/R_B \right)
 \left(
 \begin{array}{c}
   j_0\left(x_q r/R_B\right)\\
 i\beta_q \vec{\sigma} \cdot \hat r j_1\left(x_q r/R_B\right) 
 \end{array}\right)
  \frac{\chi_q}{\sqrt{4\pi}} ~,
 \end{equation}
 where 
 \begin{equation}
 \epsilon_q=\Omega_q +R_B\left(g_\omega^q\, \omega+
 \frac{1}{2} g^q_\rho \tau_z \rho_{03} \right) ~; ~~~
 \beta_q=\sqrt{\frac{\Omega_q-R_B\, m_q^*}{\Omega_q\, +R_B\, m_q^* }}\ ,
 \end{equation}
 with the normalization factor given by
 \begin{equation}
 {\cal N}_q^{-2} = 2R_B^3 j_0^2(x_q)\left[\Omega_q(\Omega_q-1)
 + R_B m_q^*/2 \right] \Big/ x_q^2 ~,
 \end{equation}
 where $\Omega_q\equiv \sqrt{x_q^2+(R_B\, m_q^*)^2}$, 
 $m_q^*=m_q^0-g_\sigma^q\, \sigma$, $R_B$ is the bag radius of the baryon, 
 and $\chi_q$ is the quark spinor. The quantities $\psi_q,\, \epsilon_q,\, 
 \beta_q,\, {\cal N}_q,\, \Omega_q,\, m^*_q$ all depend on the baryon 
 considered. The bag eigenvalue, $x_q$, is determined by the 
 boundary condition at the bag surface
 \begin{equation}
 j_0(x_q)=\beta_q\, j_1(x_q)\ .
 \label{bun-con}
 \end{equation}
 The energy of a static bag describing baryon $B$ consisting of three ground state quarks 
 can be expressed as
 \begin{equation}
 E^{\rm bag}_B=\sum_q n_q \, \frac{\Omega_q}{R_B}-\frac{Z_B}{R_B}
 +\frac{4}{3}\, \pi \, R_B^3\, B_B\ ,
 \label{ebag}
 \end{equation}
 where $Z_B$ is a parameter which accounts for zero-point motion
 and $B_B$ is the bag constant.
 The effective mass of a nucleon bag at rest is taken to be $M_B^*=E_B^{\rm bag}.$
 The equilibrium condition for the bag is obtained by 
 minimizing the effective mass, $M_B^*$ with respect to the bag radius
 \begin{equation}
 \frac{d\, M_B^*}{d\, R_B^*} = 0\ .
 \label{balance}
 \end{equation}
 For the QMC model, the equations of motion for the meson fields in
 uniform static matter are given by
 \begin{equation}
 m_\sigma^2\sigma = \sum_B g_{\sigma B} C_B(\sigma) \frac{2J_B + 1}{2\pi^2}
 \int_0^{k_B} \frac{M_B^*(\sigma)}
 {\left[k^2 + M_B^{* 2}(\sigma)\right]^{1/2}} \: k^2 \ dk ~,
 \label{field1}
 \end{equation}
 \begin{equation}
 m_\omega^2\omega_0 = \sum_B g_{\omega B} \left(2J_B + 1\right)
 k_B^3 \big/ (6\pi^2) ~,
 \label{field2}
 \end{equation}
 \begin{equation}
 m_\rho^2\rho_{03} = \sum_B g_{\rho B} I_{3B} \left(2J_B + 1\right)
 k_B^3 \big/ (6\pi^2) ~.
 \label{field3}
 \end{equation}
 In the above equations $J_B$, $I_{3B}$ and $k_B$ are respectively the spin, 
 isospin projection and the Fermi momentum of the baryon species $B$.
 For the hyperon couplings we take $x_{\omega}=0.78$ and $x_{\rho}=0.7$. The coupling $x_{\sigma}$
 is an output of the model and is approximately equal to 0.7.
 Note that the $s$-quark is unaffected by the $\sigma$ and $\omega$
 mesons i.e. $g_\sigma^s=g_\omega^s=0\ .$
 
 In Eq. (\ref{field1}) we have
 \begin{equation}
 g_{\sigma B}C_B(\sigma) = - \frac{\partial M_B^*(\sigma)}{\partial \sigma} 
 = - \frac{\partial E^{\rm bag}_B}{\partial \sigma} 
 = \sum_{q=u,d} n_q g^q_\sigma S_B(\sigma)
 \end{equation}
 where
 \begin{equation}
 S_B(\sigma) = \int_{bag} d{\bf r} \ {\overline \psi}_q \psi_q
 = \frac{\Omega_q/2 + R_Bm^*_q(\Omega_q - 1)}
 {\Omega_q(\Omega_q -1) + R_Bm_q^*/2} ~; ~~~~ q \equiv (u,d) ~.
 \end{equation}
 
 The total energy density and the pressure
 including the leptons can be obtained from the grand canonical
 potential and they read
 \begin{eqnarray}
 \varepsilon &=& \frac{1}{2}m_\sigma^2 \sigma^2
 + \frac{1}{2}m_\omega^2 \omega^2_0
 + \frac{1}{2} m_\rho^2 \rho^2_{03} \nonumber\\
 &+& \sum_B \frac{2J_B +1}{2\pi^2} \int_0^{k_B}k^2 dk
 \left[k^2 + M_B^{* 2}(\sigma)\right]^{1/2} 
 + \sum_l \frac{1}{\pi^2} \int_0^{k_l} k^2 dk\left[k^2 + m_l^2\right]^{1/2}~,
 \end{eqnarray}
 \begin{eqnarray}
 P &=& - \frac{1}{2}m_\sigma^2 \sigma^2
 + \frac{1}{2}m_\omega^2 \omega^2_0
 + \frac{1}{2} m_\rho^2 \rho^2_{03} \nonumber\\
 &+& \frac{1}{3} \sum_B \frac{2J_B +1}{2\pi^2} \int_0^{k_B}
 \frac{k^4 \ dk}{\left[k^2 + M_B^{* 2}(\sigma)\right]^{1/2}}
 + \frac{1}{3} \sum_l \frac{1}{\pi^2} \int_0^{k_l} \frac{k^4 dk}
 {\left[k^2 + m_l^2\right]^{1/2}} ~.
 \end{eqnarray}
 
 For the bag radius we take $R_N=0.6$ fm. The two unknowns $Z_N$ and $B_N$ for nucleons
 are obtained by fitting the nucleon mass $M=939$ MeV and
 enforcing the stability condition for the bag at free space. The values
 obtained are $Z_N=3.98699$ and $B_N^{1/4}=211.303$ MeV for $m_u=m_d=0$ MeV and
 $Z_N=4.00506$ and $B_N^{1/4}=210.854$ MeV for $m_u=m_d=5.5$ MeV. We take
 these bag values, $B_B$, for all baryons and the parameter $Z_B$ and $R_B$
 of the other baryons are obtained
 by reproducing their physical masses in free space and again enforcing
 the stability condition for their bags.
 Note that for a fixed bag value, the equilibrium condition in free space results in an increase of the 
 bag radius and a decrease of the parameters $Z_{B}$ for the heavier baryons.
 The set of parameters used in the present work is given in Ref. \cite{parameter}. 
 
 Next we fit the quark-meson coupling constants $g_\sigma^q$, 
 $g_\omega = 3g_\omega^q$ and $g_\rho = g_\rho^q$ for the nucleon to obtain 
 the correct saturation properties of
 the nuclear matter, $E_B \equiv \epsilon/\rho - M = -15.7$~MeV at
 $\rho~=\rho_0=~0.15$~fm$^{-3}$, $a_{sym}=32.5$ MeV, $K=257$ MeV and
 $M^*=0.774 M$. We have 
 $g_\sigma^q=5.957$, $g_{\omega N}=8.981$ and $g_{\rho N}=8.651$.
 We take the standard values for the meson masses, $m_\sigma=550$ MeV, 
 $m_\omega=783$ MeV $m_\rho=770$ MeV.

 \section{Quantum nucleation of quark matter in hadron stars}
 \label{sec:quantum}
 
 Let us consider a pure hadronic star whose central pressure (density) is increasing due 
 to spin-down or due to mass accretion (from a companion or from the interstellar medium). 
 As the central pressure approaches the deconfinement threshold pressure $P_0$ (see Fig. \ref{gm1}), 
 a drop of non $\beta$-stable quark matter ($Q^*$), but with flavor content equal to that
 of the $\beta$-stable hadronic phase, can be formed in the central region of the star. 
 The process of drop formation is regulated by its quantum fluctuations in the potential well 
 created from the difference in the Gibbs free energies of the hadron and quark phases \cite{iida97,iida98,be03}
 \begin{equation}
  U({\cal R})=\frac{4}{3}\pi n_{b,Q^*}(\mu_{Q^*}-\mu_H){\cal R}^3 + 4\pi \sigma {\cal R}^2
 \label{eq:potential}
 \end{equation}
 where ${\cal R}$ is the radius of the $Q^*$ droplet (supposed to be spherical), $n_{b,Q^*}$ is the 
 quark baryon number density, $\mu_{Q^*}$ and $\mu_H$ are the quark and hadron chemical potentials
 at a fixed pressure $P$ and $\sigma$ is the surface tension for the surface separating the hadron from the $Q^*$ phase. 
 Notice that $\mu$ is the same as the bulk Gibbs energy per baryon $g=(P+\epsilon)/n_B=(\sum_i\mu_in_i)/n_B$. Notice also that we have neglected the term associated with the curvature energy, 
 and also the terms connected with the electrostatic energy, since they are known to introduce small corrections \cite{iida98,bo04}. 
 The value of the surface tension $\sigma$ for the interface separating the quark and hadron phase 
 is poorly known, and typically values used in the literature range within $10-50$ MeV fm$^{-2}$ 
 \cite{he93,iida98}. 
 
 The time needed to form the first drop (nucleation time) can be straightforwardly evaluated within 
 a semi-classical approach \cite{iida97,iida98}. First one computes, in the Wentzel--Kramers--Brillouin (WKB) approximation, the ground state energy $E_0$ and the oscillation frequency $\nu_0$ of the drop in the potential well $U({\cal R})$. Then, the probability of tunneling is given by
 \begin{equation}
  p_0=exp\left[-\frac{A(E_0)}{\hbar}\right]
 \label{eq:prob}
 \end{equation}
 where $A$ is the action under the potential barrier which in a relativistic framework reads
 \begin{equation}
  A(E)=\frac{2}{c}\int_{{\cal R}_-}^{{\cal R}_+}\sqrt{[2{\cal M}({\cal R})c^2 +E-U({\cal R})][U({\cal R})-E]} \ ,
 \label{eq:action}
 \end{equation}
 being ${\cal R}_\pm$ the classical turning points and
 \begin{equation}
  {\cal M}({\cal R}) = \frac{4\pi}{3} \rho_H\left(1-\frac{n_{b,Q^*}}{n_H}\right)^2 {\cal R}^3
 \label{eq:mass}
 \end{equation}
 the droplet effective mass, with $\rho_H$ and $n_H$ the hadron energy density and the hadron baryon number density, respectively. The nucleation time is then equal to
 \begin{equation}
  \tau=(\nu_0 p_0 N_c)^{-1} \ ,
 \label{eq:time}
 \end{equation}
 where $N_c$ is the number of virtual centers of droplet formation in the star. A simple estimation 
 gives $N_c \sim 10^{48}$ \cite{iida97,iida98}. The uncertainty in the value of $N_c$ is expected to be within one or two orders of magnitude. In any case, all the qualitative features of our scenario will 
 not be affected by this uncertainty. 
 As a consequence of the surface effects it is necessary to have an overpressure $\Delta P= P-P_0 > 0$ with respect to the bulk transition point $P_0$ to create a drop of deconfinement quark matter in the hadronic environment. The higher the overpressure, the easier to nucleate the first drop of $Q^*$ 
 matter. In other words, the higher the mass of the metastable pure hadronic star, the shorter the time to nucleate a quark matter drop at the center of the star. 
 
 In order to explore the astrophysical implications of quark matter nucleation, following 
 ref. \cite{be03,bo04}, we introduce the concept of {\it critical mass} for the hadronic star sequence. 
 The {\it critical mass} $M_{cr}$ is the value of the gravitational mass of a metastable hadronic star for which the nucleation time is equal to one year: $M_{cr}=M_{HS} (\tau=1 yr)$. 
 Therefore, pure hadronic stars with $M_{HS} > M_{cr}$ are very unlikely to be observed, while pure
 hadronic stars with $M_{HS} < M_{cr}$ are safe with respect to a sudden transition to quark matter. 
 Then $M_{cr}$ plays the role of an {\it effective maximum mass for the hadronic branch of 
 compact stars} (see discussion in Ref.\ \cite{bo04}). 
 While the Oppenheimer--Volkov maximum mass is determined by the overall stiffness of the equation of state for hadronic matter, the value of $M_{cr}$ will depend in addition on the properties of the intermediate non $\beta$-stable $Q^*$ phase.

 \section{Results and discussion}
 \label{sec:results}
 
 In this section we present and discuss our results for stellar configurations obtained using 
 the equation of state (EoS) models described in section II. 
 In particular, we determine the region of the pure hadronic star sequence where these compact stars are metastable, 
 the value of the corresponding critical mass $M_{cr}$, and the final fate of this configuration after quark matter 
 nucleation, {\it i.e.} whether it will evolve to a quark star or to a black hole. 
 
\begin{figure}[t]
 \begin{tabular}{cc}
 \includegraphics[height=12cm,angle=-90]{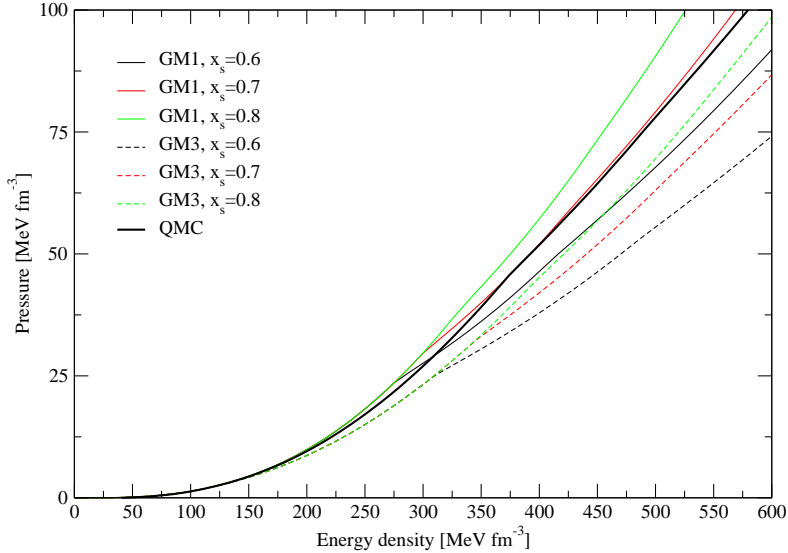} &
 \end{tabular}
 \caption{Hadronic Eos for QMC and for GM1 and GM3 with the different hyperon-meson
   couplings discussed in the text.}
 \label{eos}
 \end{figure}

 In Fig \ref{eos} the EoS for the models discussed are plotted for the range of densities 
 of relevance for the discussion that follows.
 For GM1 and GM3 we have considered three different hyperon-meson coupling as 
 discussed above. The QMC EoS corresponds approximately to $x_{\sigma}=0.7$.
 A higher value of the hyperon couplings $x_i$ corresponds to stiffer EoSs: at
 high densities we have vector dominance defined by the magnitude of $x_{\omega}, \,
 x_{\rho}$. It is clear from Fig. \ref{eos} that the onset of hyperons 
 (represented by the change of slope in the EoS curves) 
 occurs for the smaller $x_{\sigma}$ values at lower densities. 
 The nucleonic EoS for QMC is very soft and therefore the onset of hyperons occurs at quite high 
 densities, $\varepsilon=373.87$ MeV/fm$^3$. As a consequence although QMC is softer than 
 GM1 EoS at lower densities, it becomes, at higher densities, stiffer than GM1($x_{\sigma}=0.6$) 
 and very close to GM1($x_{\sigma}=0.7$).
 
\begin{figure}[t]
 \begin{tabular}{c}
 \includegraphics[height=10cm,angle=-90]{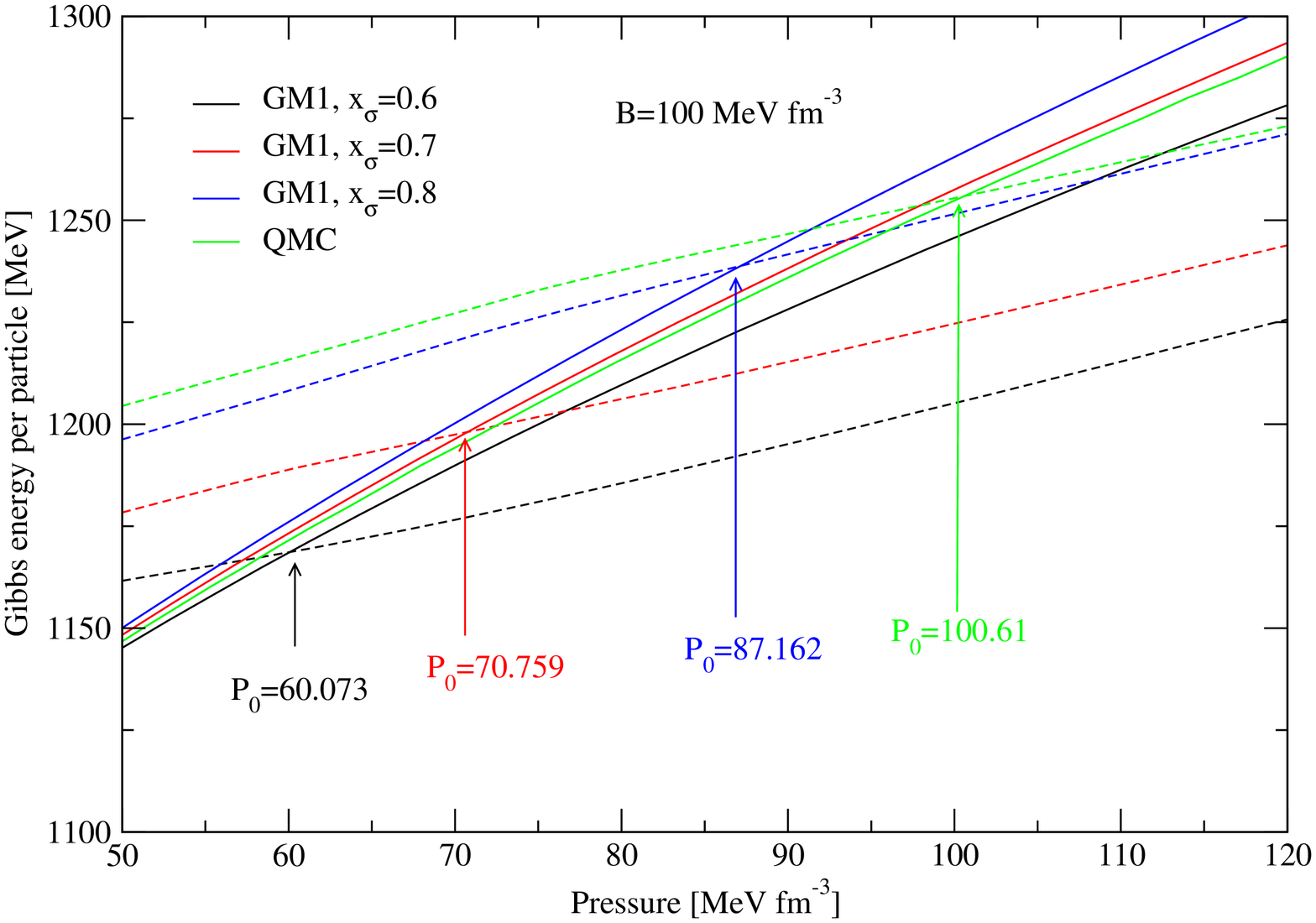} \\
 \includegraphics[height=10cm,angle=-90]{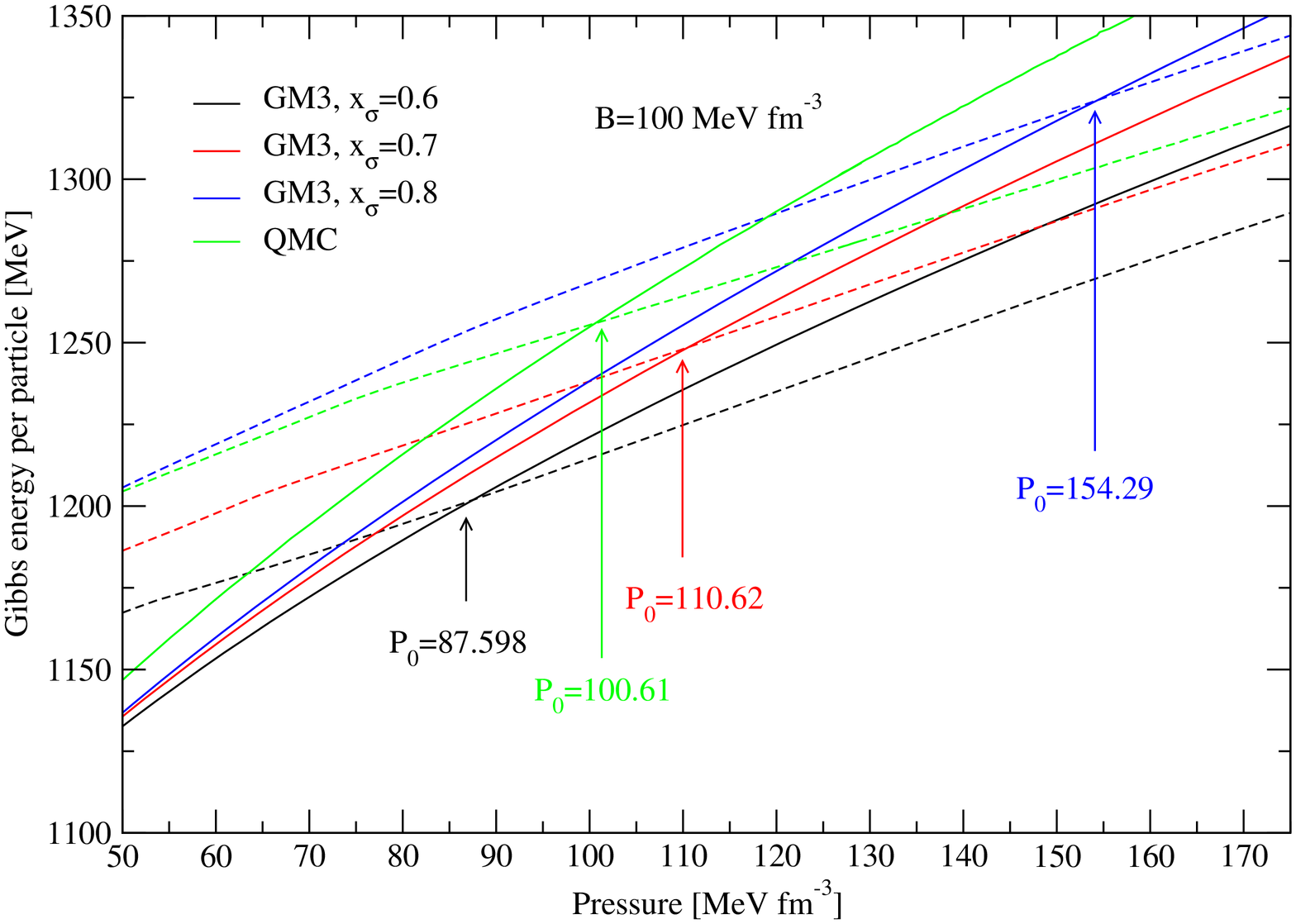} 
 \end{tabular}
 \caption{The Gibbs` energy per particle for the $\beta$-stable hadronic phase (continuous curves) 
 and for the respective $Q^*$ phase (dashed curves). The upper panel refers to the 
 GM1 and the lower panel to the GM3 EoS. The results for the QMC model are plotted in both panels.}
 \label{gm1}
 \end{figure}

 In Figs. \ref{gm1}, we plot the Gibbs` free energy per baryon for the hadronic phase and for 
 the corresponding $Q^*$ phase using the various EoS models (couple of continuous and dashed 
 curves with the same color) considered in the present work. 
 It is clearly seen that in the case of the GM1 or GM3 EoS models, the lower the value 
 of the hyperon coupling $x_{\sigma}$, the softer the EoS (see also Fig \ref{eos}) and the lower 
 the pressure $P_0$ at the crossing between the hadronic and the $Q^*$ phase. 
 This will give rise to lower critical masses for the smaller $x_{\sigma}$ values (see Tabl. I-III below). 
 The $Q^*$ phase is very sensitive to the particle content and it is due to this fact that, 
 although in Fig. \ref{gm1} the EoS for QMC is softer than the EoS for GM1 with $x_{\sigma}=0.8$ and 0.7, 
 its crossing with the $Q^*$ phase occurs at higher pressures. 
 A similar observation occurs in the figure with the GM3 results. 
 This behavior will reflect itself on values of the critical masses $M_{cr}$. 
 
\begin{figure}[t]
 \begin{tabular}{cc}
 \includegraphics[height=9cm,angle=0]{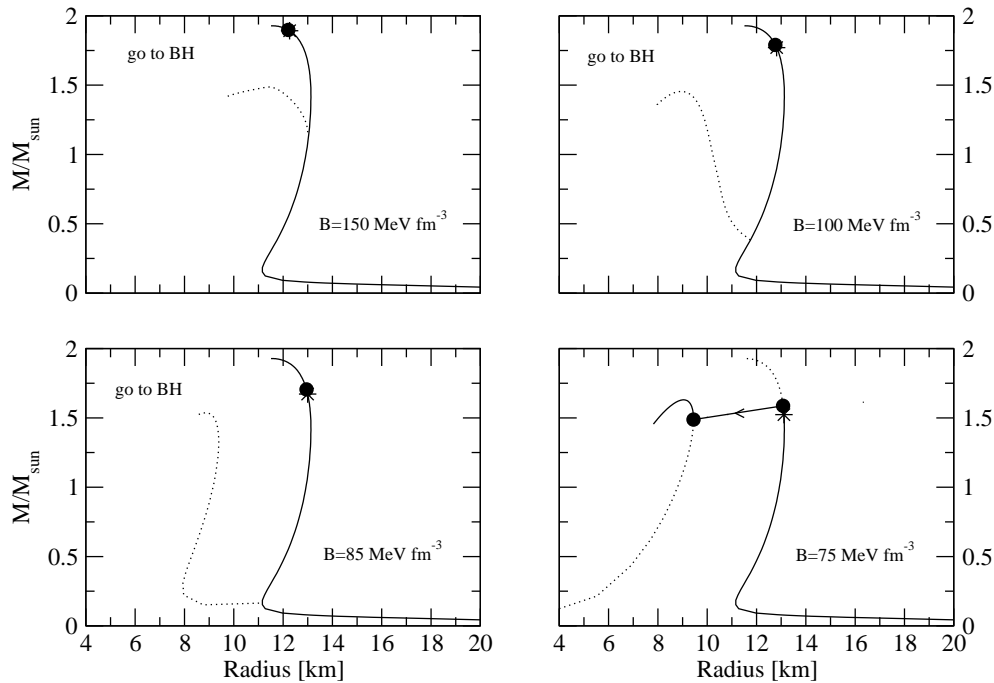} \\
 \end{tabular}
 \vspace{0.8cm}
 
 \caption{Mass-radius relation for a pure HS described within the QMC model 
 and that of the HyS or SS configurations for several values of the bag constant and 
 $m_s=150$ MeV and $\alpha_s=0$. The configuration marked with an asterisk represents 
 in all cases the HS for which the central pressure is equal to $P_0$. 
 The conversion process of the HS, with a gravitational mass equal to $M_{cr}$, into
 a final HyS or SS is denoted by the full circles connected by an arrow. 
 In all the panels $\sigma$ is taken equal to $30$ MeV/fm$^2$.}
 \label{QMC_MR}
 \end{figure}

 In Fig.\ \ref{QMC_MR}, we show the mass-radius (MR) curve for pure HS within the QMC model
 for the EoS of the hadronic phase, and that for hybrid stars 
 or strange stars for different values of the bag constant $B$. 
 The configuration marked with an asterisk on the hadronic MR curves represents 
 the hadronic star for which the central pressure is equal to the threshold value $P_0$ 
 and the quark matter nucleation time is $\tau = \infty$. 
 The full circle on the hadronic star sequence represents the critical mass configuration, 
 in the case $\sigma = 30$ MeV/fm$^2$. 
 The full circle on the HyS (SS) mass-radius curve represents the hybrid (strange) star 
 which is formed from the conversion of the hadronic star with $M_{HS} = M_{cr}$. 
 We assume \cite{grb} that during the stellar conversion process the total number of baryons 
 in the star (or in other words the stellar baryonic mass) is conserved. Thus the total energy 
 liberated in the stellar conversion is given by the difference between the gravitational mass of 
 the initial hadronic star ($M_{in} \equiv M_{cr}$) and that of the final hybrid or strange stellar configuration with the same baryonic mass ($M_{fin} \equiv M_{QS}(M^b_{cr}) \,$): 
 \begin{equation}
           E_{conv} = (M_{in} - M_{fin}) c^2 \, . 
 \label{eq:eq11}
 \end{equation}
 As we can see from Fig.\ \ref{QMC_MR}, for the case of the QMC model, the region of metastability 
 of pure hadronic stars (the part of the MR curve between the asterisk and the full circle) 
 is very narrow. For this hadronic EoS, the quark star sequence can be populated only in the case of 
 ``small'' values of the bag constant ($B \leq 80$ MeV/fm$^3$, in this case the final star is a strange star). 
 In all the other cases the critical mass hadronic star will form a black hole.

 For comparison we plot the MR curve obtained with the GM1 parametrization for
 the same surface tension ($\sigma =30 $MeV/fm$^2$) and two values for bag constant 
 ($B=75$ and 100 MeV/fm$^3$). We consider the two extreme values of the hyperon 
 couplings studied in this work. The dots and stars have the same meaning as in
 Fig.\ \ref{QMC_MR}. We see that for the cases plotted the only configuration
 that does not end in a black hole has the smallest bag constant and hyperon
 coupling considered. In the present model, however, the configuration with the
 central pressure $P_0$ and the $M_{cr}$ configuration are quite separated, contrary 
 to what was observed with QMC, Fig.\ \ref{QMC_MR}.
 
\begin{figure}[t]
 \begin{tabular}{cc}
 \includegraphics[height=9cm,angle=0]{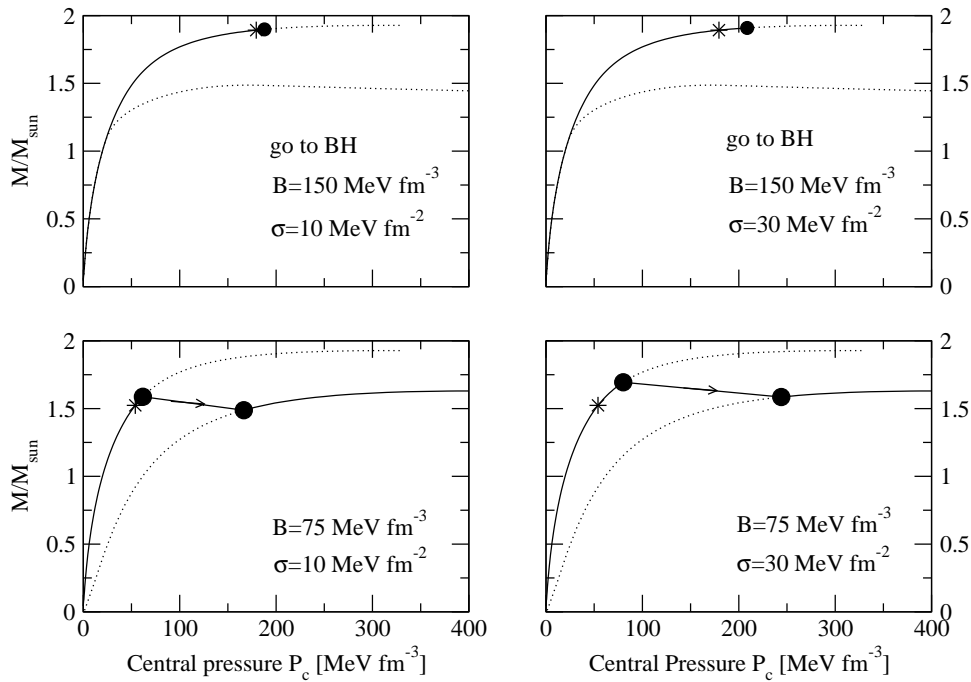} 
 \end{tabular}
 \vspace{0.3cm}
 
 \caption{Mass-pressure relation for a pure HS described within the QMC model and 
 that of the HyS or SS configurations two values of the Bag constant(75 and 100
 MeV/fm$^3$) and 
 $m_s=150$ MeV and $\alpha_s=0$. The configuration marked with an asterisk represents 
 in all cases the HS for which the central pressure is equal to $P_0$. 
 The conversion process of the HS, with a gravitational mass equal to $M_{cr}$, into
 a final HyS or SS is denoted by the full circles connected by an arrow. 
 Two values of the surface energy $\sigma$ were considered $10$ MeV/fm$^2$
 (left) and $30$ MeV/fm$^2$ (right).}
 \label{QMC_MP}
 \end{figure}

\begin{figure}[t]
 \begin{tabular}{cc}
\includegraphics[height=9cm,angle=0]{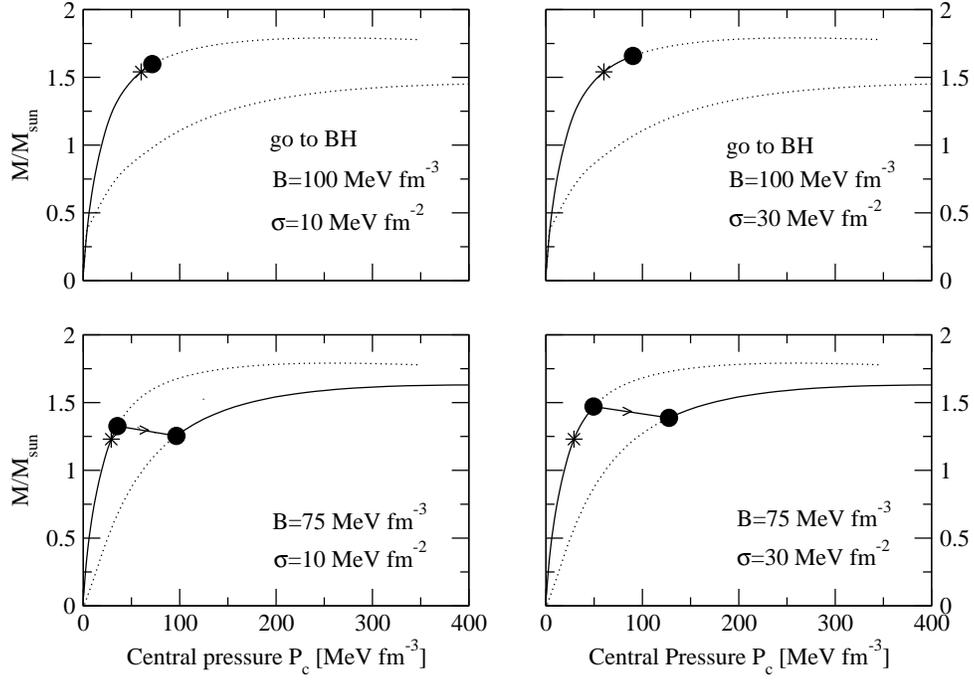} 
\end{tabular}
\vspace{0.3 cm}
 
 \caption{The same as Fig \ref{QMC_MP} for the GM1
  parametrization with the $x_\sigma=0.6$ hyperon coupling.
 }
 \label{GM1_MP}
 \end{figure}

 The larger mass difference between the star with the central pressure $P_0$
 and the one with the $M_{cr}$ occurs when these stars have small masses.
 A small change in the central energy density corresponds
 to a large change in the mass. If instead of plotting the MR graph we
 would have plotted the corresponding mass--central pressure (MP) graph a larger
 difference between these two configurations would be expected. This is seen in
 Figs. \ref{QMC_MP} and \ref{GM1_MP} where the mass-pressure curves for the
 family of stars obtained respectively within QMC and GM1 are plotted for two
 bag constants and two values of the surface tension ($\sigma = 10$ and 30 MeV/fm$^2$). 
 We conclude that when the $M_{cr}$ star is almost on top of the $P_0$ star in the MR curves,
 these stars lie on or close to the plateau that contains the
 maximum mass configuration. A large separation between these two configurations
 corresponds to a phase transition which occurs during the rise of the
 MR curve before the plateau. Due to the softness of the QMC EOS,
 hyperons set on at quite large energy densities and
 the star with the central $P_0$ pressure only occurs at high densities.
 We also conclude that a smaller surface tension hastens the transition 
 and the critical mass is closer to the $P_0$ mass.

 \begin{figure}[t]
 \begin{tabular}{cc}
 \includegraphics[height=9cm,angle=0]{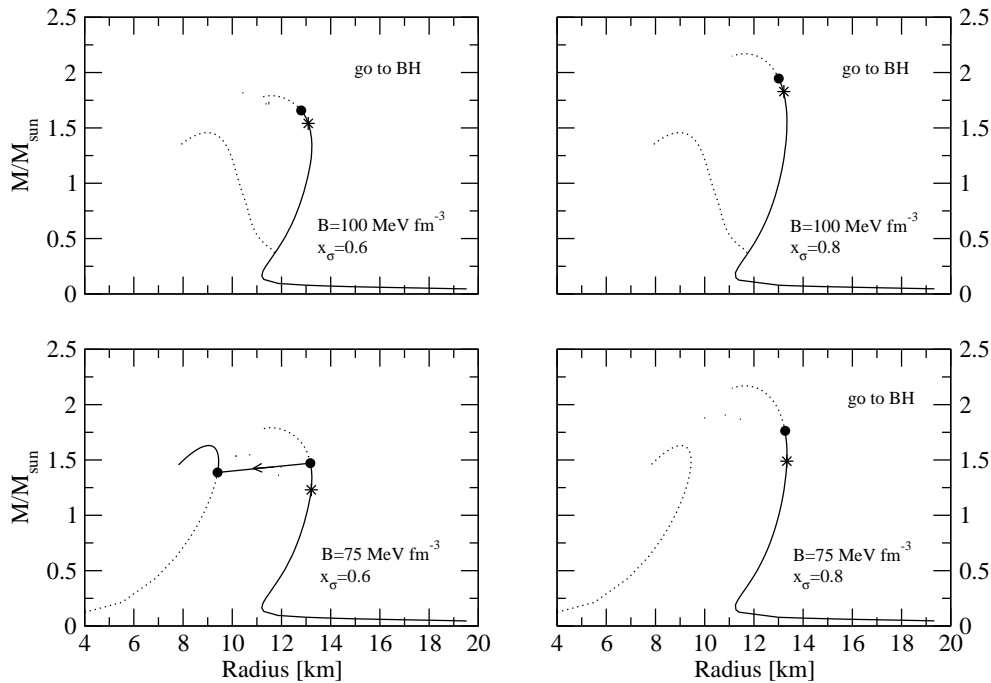} \\
 \end{tabular}
 \vspace{0.8cm}
 
 \caption{
 Mass-radius relation for a pure HS described within the GM1 parametrization 
 and that of the HyS or SS configurations for two values of the bag constant
 ($B=75$ and 100 MeV/fm$^3$) and two values of the hyperon-meson coupling 
 ($x_\sigma=0.6,$ and 0.8) and $m_s=150$ MeV and $\alpha_s=0$. 
 The configuration marked with an asterisk represents 
 in all cases the HS for which the central pressure is equal to $P_0$. 
 The conversion process of the HS, with a gravitational mass equal to $M_{cr}$, into
 a final HyS or SS is denoted by the full circles connected by an arrow. 
 In all the panels $\sigma$ is taken equal to 30 MeV/fm$^2$.}
 \label{GM1_MR}
 \end{figure}

 In Tables \ref{tabgm1}, \ref{tabgm3}, and \ref{tabqmc} 
 we give the gravitational ($M_{cr}$) and baryonic ($M_{cr}^b$) critical mass values 
 for the hadronic star sequence, together with the central hyperon fraction ($f_{Y,cr} = n_Y/n_B$, 
 {\it i.e.} the ratio between the total hyperon number density and the total baryon number density 
 at the center of the critical mass star). 
 We also report the value of the gravitational mass ($M_{fin}$) 
 of the final quark star configuration and the total energy \cite{grb} 
 $E_{conv} = (M_{cr} - M_{fin}) c^2$ released in the stellar conversion process, assuming 
 baryon mass conservation ({\it i.e.} no matter ejection) \cite{grb}. 
 The gravitational ($M_{QS,max}$) and the baryonic ($M_{QS,max}^b$) mass 
 of the maximum mass configuration for the quark (hybrid or strange) star
 sequence are also included. 
 The value of the latter quantity is relevant to establish whether the critical mass 
 hadronic star will evolve to a quark star ($M_{cr}^b < M_{QS,max}^b$) or will form a 
 black hole ($M_{cr}^b > M_{QS,max}^b$). 
 The entries in Tables \ref{tabgm1}, \ref{tabgm3}
 {\footnote{We have found a typo in one of the GM EoS coupling constants in the code used by 
 the authors of ref. \cite{bo04}. In the present calculations, we have corrected this typo and 
 we have increased the numerical accuracy of our code. This justifies the small differences 
 between the present results and those reported in ref. \cite{bo04}.}}, 
 and \ref{tabqmc} are relative respectively 
 to the GM1, GM3 and QMC equation of state for the hadronic phase. 
 For the quark phase we consider four different values of bag constants, 75, 85, 100 and 
150 MeV/fm$^3$,  and two different values for quark-hadron surface tension, 10 and 30 MeV. 
Notice that for the quark matter parameter set adopted in the present work (see Section I), 
strange quark matter is  absolutely stable \cite{bod71,witt84} only for $B = 75$ MeV/fm$^3$.  

Some comments are in order: the critical masses increase with the increase of the hyperon couplings.  
This increase can be as large as 0.3 - 0.4 $M_\odot$ when $x_{\sigma}$ changes from 0.6 to 0.8;   
the critical mass is also dependent on the particle content, namely of the strangeness content,   
and this explains the different relative positions for the different bag pressures of the  
QMC result which essentially corresponds to $x_{\sigma}=0.7$. Due to the fact that the EoS for QMC  
is very soft, the hyperon onset occurs at quite high densities and therefore the critical mass is  
always quite high for this model.   
The critical mass increases with the bag constant because a larger bag constant corresponds to a  stiffer quark EoS and therefore the phase transition to the quark phase will occur at   
larger densities.  When the critical mass hadronic star is converted to a black hole, this is 
indicated in Tables \ref{tabgm1}, \ref{tabgm3}, and \ref{tabqmc} with a entry BH, in the columns  
for $M_{fin}$ and $E_{conv}$ (no energy will be radiated as soon as the star pass the event horizon). 
Notice that, in the case of the GM3 model with $x_{\sigma}=0.6$ and $B = 150$ MeV/fm$^3$ 
there is no entry for the critical mass value (and for $M_{cr}^b$, $M_{fin}$ and $E_{conv}$) since in this case the nucleation time of the maximum mass hadronic star ($M_{HS,max}$) is much larger 
than one year ({\it i.e.} the star is metastable with a {\it life-time} comparable or much higher 
than the age of the universe). 
 
  We observe from the results in tables \ref{tabgm1}, \ref{tabgm3}, that increasing the value 
 of the hyperon coupling constants (for fixed $B$ and $\sigma$) reduces the central hyperon fraction 
 ($f_{Y,cr}$) of the critical mass star, and increases the energy released during the conversion into 
 a quark or hybrid star (for those configurations which will not form a black hole).  

\begin{figure}[t]
 \begin{tabular}{cc}
\includegraphics[height=8.5cm,angle=0]{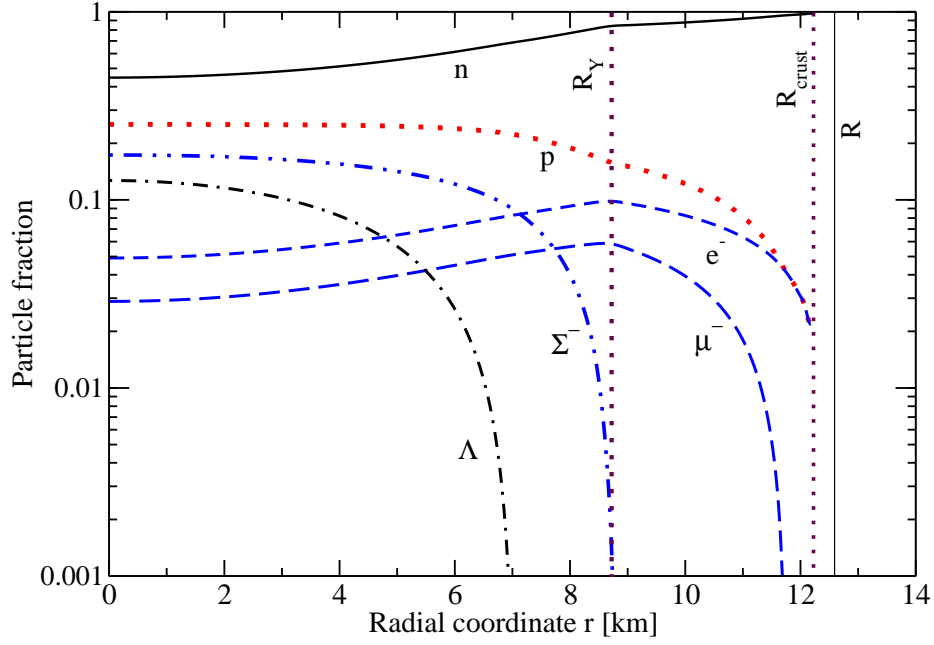} 
\end{tabular}
\vspace{0.3 cm}
 
\caption{The internal composition for an hadronic star with gravitational mass $M = 2.081 M_\odot$ and radius $R = 12.6$ km, obtained with the GM1 equation of state with $x_{\sigma} = 0.8$.  
This star corresponds to the critical mass configuration when we consider $B = 150$ MeV/fm$^3$ 
and $\sigma = 30$ MeV/fm$^2$  (see table \ref{tabgm1}). 
$R_Y$ is the radius of the hyperonic matter stellar core. The nuclear matter layer extends between $R_Y$ and $R_{crust}$. The crust extends between  $R_{crust}$ and  $R$. } 
\label{profile}
\end{figure}

In Fig. \ref{profile}, we show the internal composition for the hadronic star with a gravitational 
mass  $M= 2.081 M_\odot$ and radius $R = 12.6$ km, obtained using the GM1 parametrization 
with $x_{\sigma} = 0.8$.  This star corresponds to the critical mass configuration when we 
consider  $B = 150$ MeV/fm$^3$ and $\sigma = 30$ MeV/fm$^2$  (see table \ref{tabgm1}). 
As we see, this star has a considerable central hyperon fraction  ($f_{Y,cr} = 0.299$) and a wide 
hyperonic matter core which extend up to $R_Y \sim 8.7$ km. On the top of this core, one has a nuclear matter layer ($R_Y \leq r \leq R_{crust}$) with a thickness of about 3.4 km. 
The stellar crust extends from $R_{crust}$ up to $R$.

It has been argued by several authors \cite{afo86,mad88,ob89,cf91} that if strange quark 
matter (SQM) is absolutely stable \cite{bod71,witt84}, then all compact stars are likely to 
be strange stars.  
The argument in favor of this thesis is the following: if the interstellar medium is sufficiently  
contaminated by {\it quark nuggets} ({\it i.e.} lumps of SQM), then the presence of a single quark nugget in the interior of a ``normal'' neutron star (hadronic star)  is sufficient to trigger the conversion of the star to a strange star \cite{witt84,afo86}.   
Likely the quark nugget contamination of the interstellar medium is the result of the merging of strange stars in binary systems \cite{mad88,mad06}. 
Under these conditions, compact star progenitors could capture a quark nugget during their {\it lives} ({\it i.e.} during the various nuclear burning stages of the stellar evolution). Thus, according to this argument, a strange quark seed will be present in all new born compact stars, and thus the conversion to a strange star will happen immediately, without a metastable hadronic star being formed first.  
This is a plausible scenario, however it would be relevant only for a few of the stellar models    considered in our work, {\it i.e.} those relative to the value $B = 75$ MeV/fm$^3$ for the bag   constant (see Tables \ref{tabgm1}, \ref{tabgm3}, and \ref{tabqmc}) for which SQM is absolutely  
stable, and will not have any effect upon the existence of metastable hadronic compact stars  
in all the other cases considered in the present work.              
The magnitude of the flux of quark nuggets in the interstellar  medium (which is  
a crucial quantity for the validity of the scenario of ref.s \cite{afo86,mad88,ob89,cf91}) 
has been estimated \cite{mad88} making the assumption that all pulsars exhibiting glitches must be ``normal'' neutron stars (hadronic stars), not strange stars. This assumption is based primarily on the nearly total lack of models for the glitch phenomenon with strange stars (see anyhow ref. \cite{horv04,xu}), while such models have been quite successfully developed in the case of hadronic 
stars (see {\it e.g.} ref. \cite{ai75,eb88,leb93,jon97}).  
However, recent studies have established the possibility of an inhomogeneous crystalline color superconducting phase (LOFF phase) in the interior of strange stars (see \cite{mrs07,cn04} and 
references therein quoted), or the likely existence of a SQM crystalline crust in strange stars 
\cite{jrs06}.  These late theoretical developments rise the possibility to explain pulsar  
glitches with strange star based models, and thus require as useful a recalculation of the 
astrophysical limits of the flux of quark nuggets.  
The scenario discussed in the present work is an alternative to the 
scenario \cite{afo86,mad88,ob89,cf91} according to which all compact stars are strange stars, 
which requires that SQM is absolutely stable.    
 
\section{Conclusions}
\label{sec:conclusions} 
 It has been recently shown \cite{be03,bo04,drago04,lug05,blv07} that pure hadronic compact stars, above a threshold value of their gravitational mass, are metastable to the conversion to quark stars.   
In this work we have done a systematic study of the metastability of pure hadronic compact stars 
 using different relativistic hadronic models for the equation of state of hadronic dense matter. 
 In particular, we have used and compared the quark-meson coupling (QMC) model with those for the Glendenning--Moszkowski parametrization of the non-linear Walecka model (NLWM). 
In the case of the QMC model, we have obtained that the region of metastability of pure 
hadronic stars is very narrow. 
For the GM model, we have investigated the effect of the hyperon couplings on the critical mass  
of the hadronic star sequence and on the stellar conversion energy. We have found that increasing  
the value of the hyperon coupling constants shifts the bulk transition point for quark deconfinement  
to higher densities, increasing the value of the critical mass for the hadronic stellar sequence,  
and thus makes the formation of quark stars less likely. 
The nucleonic EoS for QMC is very soft and therefore the onset of hyperons occurs at quite high densities, which gives rise to large critical masses. The conversion to a quark star will occur 
only for a small value of the bag constant. Finally we point out that both QMC and GM1 with 
the largest values of the hyperon-meson couplings predict {\it limiting masses} \cite{bo04}  
which may be as high as 1.9 - 2.1 $M_\odot$. 

These values would be able to describe highly-massive compact stars, such as the one associated 
to the millisecond pulsars PSR~B1516+02B \cite{frei07a}, and nearly the one in 
PSR~J1748-2021B \cite{frei07b}.

 \begin{table}[ht]
 \caption{Critical masses and energy released in the conversion process of an HS into a QS, 
 for several values of the hyperon coupling $x_{\sigma}$, of the Bag constant $B$ and the surface 
 tension $\sigma$. The GM1 parameter set has been used for the hadronic EoS. 
 Column labeled $M_{QS,max}$ ($M^b_{QS,max}$) denotes the maximum gravitational 
 (baryonic) mass of the final QS sequence. 
 The value of the critical gravitational (baryonic) mass of the initial HS is reported on 
 column labeled $M_{cr}$ ($M^b_{cr}$) whereas those of the mass of the final QS and the 
 energy released in the stellar conversion process are shown on columns labeled 
 $M_{fin}$ and $E_{conv}$ respectively. 
 BH denotes those cases in which due to the conversion the initial HS collapses into a black hole.
 $f_{Y,cr} = n_Y/n_B$ denotes central hyperon fraction of the critical mass star ({\it i.e.} the ratio between the total hyperon number density and the total baryon number density at the center of the critical mass star). 
 Units of B and $\sigma$ are MeV/fm$^3$ and MeV/fm$^2$ respectively. 
 All masses are given in solar mass units and the energy released is given in units of 
 $10^{51}$ erg. $m_s$ and $\alpha_s$ are always taken equal to $150$ MeV and $0$ respectively.  
The Oppenheimer--Volkoff  maximum masses for pure hadronic stars in the case of 
the GM1 EoS are:   
$M_{HS,max} = 1.790 M_\odot$ (when $x_\sigma = 0.6$),  1.996 $M_\odot$ ($x_\sigma = 0.7$), 
and  2.169 $M_\odot$ ($x_\sigma = 0.8$).    
}  

 \vspace{.3 cm}
 \begin{tabular}{ccccccccccccccc}\hline\hline
 & & & & \multicolumn{4}{c}{$\sigma=10$} && \multicolumn{4}{c}{$\sigma=30$} \\
 \cline{5-8} \cline{10-13} \\
 \multicolumn{1}{c}{$x_{\sigma}$} &
 \multicolumn{1}{c}{B} 
 & \multicolumn{1}{c}{$M_{QS,max}$} &
 \multicolumn{1}{c}{$M^b_{QS,max}$} & \multicolumn{1}{c}{$M_{cr}$} &
 \multicolumn{1}{c}{$M^b_{cr}$} & \multicolumn{1}{c}{$f_{Y,cr}$} &
 \multicolumn{1}{c}{$M_{fin}$} & \multicolumn{1}{c}{$E_{conv}$} & &
 \multicolumn{1}{c}{$M_{cr}$} &
 \multicolumn{1}{c}{$M^b_{cr}$} & \multicolumn{1}{c}{$f_{Y,cr}$} &
 \multicolumn{1}{c}{$M_{fin}$} & \multicolumn{1}{c}{$E_{conv}$} \\
 \hline \\
 0.6 & 75 &1.630 & 1.968 & 1.326 & 1.454 & 0.079 & 1.254 & 128.4 && 1.471& 1.630& 0.147 & 1.387 & 149.4\\
       & 85 &1.542 & 1.812 & 1.447 & 1.596 & 0.134 & 1.385 & 110.4 && 1.540& 1.711& 0.201 & 1.479 & 125.5\\
       &100&1.457 & 1.661 & 1.598 & 1.789 & 0.261 & BH & BH && 1.658& 1.865& 0.332 & BH & BH \\
       &150&1.447 & 1.601 & 1.770 & 2.010 & 0.527 & BH & BH & & 1.790& 2.036& 0.637 & BH & BH \\
 \hline\\ 
 0.7 & 75 & 1.630 & 1.968 &1.442 & 1.595 &0.059 &1.361 & 144.8& & 1.602 &1.794 &0.119&1.507 &169.6\\
      & 85 & 1.542& 1.812 & 1.584 & 1.764 &0.111 & 1.509 & 134.2 & & 1.685 &1.892 &0.170 & BH& BH\\
      &100& 1.457& 1.661 & 1.724 & 1.950 &0.197 &BH & BH & & 1.791 & 2.036 & 0.253& BH& BH\\
      &150& 1.518& 1.686 & 1.905 & 2.188 & 0.370&BH & BH & &1.931 & 2.223 & 0.403 &BH & BH\\
 \hline\\
 0.8 & 75& 1.630& 1.968 & 1.592 & 1.782 & 0.044& 1.498 & 167.9 && 1.763 & 2.000 &0.095 & BH & BH \\
      & 85 & 1.542 & 1.812 & 1.735 & 1.954 & 0.085& BH & BH && 1.841 & 2.091 &0.132 & BH & BH \\
      &100 & 1.457 & 1.661 & 1.879 & 2.152 & 0.152& BH & BH &&1.946 & 2.243 & 0.194 & BH & BH \\
      & 150 & 1.518& 1.686 & 2.054 & 2.391 & 0.275& BH & BH && 2.081 & 2.429 & 0.299& BH & BH \\
 \hline
 \end{tabular}
 \label{tabgm1}
 \end{table}
 
 \begin{table}[ht]
 \caption{Same as Tab. I, but for the GM3 parameter set for the hadronic EoS.
The Oppenheimer--Volkoff maximum masses for pure hadronic stars in the case of the GM3 EoS are:   
$M_{HS,max} =  1.554 M_\odot$ (when $x_\sigma = 0.6$),  1.732 $M_\odot$ ($x_\sigma = 0.7$), 
and 1.875 $M_\odot$ ($x_\sigma = 0.8$). }
 \vspace{.3 cm}
 \begin{tabular}{ccccccccccccccc}\hline\hline
 & & & & \multicolumn{4}{c}{$\sigma=10$} && \multicolumn{4}{c}{$\sigma=30$} \\
 \cline{5-8} \cline{10-13} \\
 \multicolumn{1}{c}{$x_{\sigma}$} &
 \multicolumn{1}{c}{B} 
 & \multicolumn{1}{c}{$M_{QS,max}$} &
 \multicolumn{1}{c}{$M^b_{QS,max}$} & \multicolumn{1}{c}{$M_{cr}$} &
 \multicolumn{1}{c}{$M^b_{cr}$} & \multicolumn{1}{c}{$f_{Y,cr}$} &
 \multicolumn{1}{c}{$M_{fin}$} & \multicolumn{1}{c}{$E_{conv}$} & &
 \multicolumn{1}{c}{$M_{cr}$} &
 \multicolumn{1}{c}{$M^b_{cr}$} & \multicolumn{1}{c}{$f_{Y,cr}$} &
 \multicolumn{1}{c}{$M_{fin}$} & \multicolumn{1}{c}{$E_{conv}$} \\
 \hline \\
 0.6 & 75 & 1.630& 1.968 & 1.237 & 1.351 & 0.092 & 1.175 & 111.6 && 1.269 & 1.389 & 0.110 & 1.204 & 115.4\\
        & 85 & 1.543& 1.812 & 1.350 & 1.482 & 0.165 & 1.298 & 91.3 && 1.362 & 1.497 & 0.178 & 1.310 & 92.8\\
        &100 & 1.465& 1.673 & 1.461 & 1.626 & 0.307 & 1.431 & 54.9 && 1.469 & 1.636 & 0.316 & 1.438 & 55.8\\
        &150 & 1.487& 1.658 & --- & --- & --- & --- & --- && --- & --- & --- & --- & --- \\
 \hline \\
 0.7 & 75 & 1.630 & 1.968 & 1.373 & 1.510 & 0.078 & 1.297 & 136.4 && 1.402 & 1.545 & 0.091 &1.324 & 140.4\\
       & 85 & 1.543 & 1.812 & 1.511 & 1.680 & 0.157 & 1.447 & 113.2 && 1.541 & 1.717 & 0.182 & 1.475 & 117.6\\
      &100 & 1.465 & 1.673 & 1.610 & 1.806 & 0.253 & BH & BH && 1.645 & 1.851 & 0.295 & BH & BH \\
      &150 & 1.495 & 1.667 & 1.716 & 1.945 & 0.419 & BH & BH && 1.723 & 1.956 & 0.444 & BH & BH \\
 \hline \\
 0.8 & 75 & 1.630 & 1.968 & 1.574 & 1.759 & 0.076 & 1.482 & 165.2 && 1.611 & 1.806 & 0.092 & 1.516 & 170.7\\
      & 85 & 1.543 & 1.812 & 1.694 & 1.913 & 0.143 & BH & BH && 1.744 & 1.979 & 0.181 & BH & BH \\
      &100 & 1.465 & 1.673 & 1.771 & 2.014 & 0.204 & BH & BH && 1.802 & 2.057 & 0.234 & BH & BH \\
      &150 & 1.495 & 1.668 & 1.848 & 2.119 & 0.295 & BH & BH && 1.856 & 2.131 & 0.309 & BH & BH \\
 \hline
 \end{tabular}
 \label{tabgm3}
 \end{table}
 
 \vspace{.3 cm}
 
 \begin{table}[ht]
 \caption{Same as Tab. I, but for the QMC parameter set for the hadronic EoS. 
The Oppenheimer--Volkoff maximum mass for pure hadronic stars in the case of the QMC EoS is:   
$M_{HS,max} =  1.927 M_\odot$. }
 \vspace{.3 cm}
 \begin{tabular}{cccccccccccccc}\hline\hline
 & & & \multicolumn{4}{c}{$\sigma=10$} && \multicolumn{4}{c}{$\sigma=30$} \\
 \cline{5-8} \cline{10-13} \\
 \multicolumn{1}{c}{B} 
 & \multicolumn{1}{c}{$M_{QS,max}$} &
 \multicolumn{1}{c}{$M^b_{QS,max}$} & \multicolumn{1}{c}{$M_{cr}$} &
 \multicolumn{1}{c}{$M^b_{cr}$} & \multicolumn{1}{c}{$f_{Y,cr}$} &
 \multicolumn{1}{c}{$M_{fin}$} & \multicolumn{1}{c}{$E_{conv}$} & &
 \multicolumn{1}{c}{$M_{cr}$} &
 \multicolumn{1}{c}{$M^b_{cr}$} & \multicolumn{1}{c}{$f_{Y,cr}$} &
 \multicolumn{1}{c}{$M_{fin}$} & \multicolumn{1}{c}{$E_{conv}$} \\
 \hline \\
  75 & 1.630 & 1.968 & 1.587 & 1.768 & 0.044&1.488 &176.8 && 1.694 & 1.903 & 0.090 & 1.585 & 195.8\\
  85 & 1.530 & 1.793 & 1.705 & 1.917 & 0.096& BH & BH && 1.768 & 1.998 & 0.145 & BH & BH \\
 100 &1.454 & 1.656 & 1.790 & 2.027 & 0.168& BH & BH && 1.830 & 2.080 & 0.211 & BH & BH \\
 150 &1.479 & 1.638 & 1.898 & 2.171 & 0.352& BH & BH && 1.909 & 2.187 & 0.377 & BH & BH \\
 \hline
 \end{tabular}
 \label{tabqmc}
 \end{table}
 
 
 \section*{Acknowledgments}
This work was partially supported by FEDER/FCT (Portugal) under the projects POCI/FP/63918/2005 
and PTDC/FIS/64707/2006 and by the Ministero dell'Universit\`a e della Ricerca (Italy) under the 
PRIN 2005 project {\it Theory of Nuclear Structure and Nuclear Matter}.

 \end{document}